\documentclass[aps,prl,twocolumn,superscriptaddress,showpacs]{revtex4} 
\newcommand{\figsize}{0.48\textwidth}      
\newcommand{\figsizes}{0.48\textwidth}      
\usepackage{graphicx}
\usepackage{dcolumn}
\usepackage{bm}
\usepackage{setspace}
\def\met{\mbox{${\hbox{$E$\kern-0.6em\lower-.1ex\hbox{/}}}_T$}} 
\def\mex{\mbox{${\hbox{$E$\kern-0.6em\lower-.1ex\hbox{/}}}_x$}} 
\def\mey{\mbox{${\hbox{$E$\kern-0.6em\lower-.1ex\hbox{/}}}_y$}} 
\def\mexy{\mbox{${\hbox{$E$\kern-0.6em\lower-.1ex\hbox{/}}}_{x,y}$}} 
 
\def\D0{D\O}                            
\newcommand{\Dzero}{\rm{D0}}

\newcommand{\etjet}{E_{  T}^{\rm jet}}
\newcommand{\etjetem}{E_{  T,{\rm em}}^{\rm jet}}
\newcommand{\etjeta}{E_{ T}^{\rm jet(1)}}
\newcommand{\etjetb}{E_{ T}^{\rm jet(2)}}

\newcommand{\etjeti}{E_{ T}^{\rm jet(i)}}
\newcommand{\etajet}{\eta^{\rm jet}}

\newcommand{\pttrk}{p_{T,\rm track}}

\def\tt{\mbox{$t \overline{t}$}}

\newcommand{\sbo}{\widetilde{b}_1}
\newcommand{\neu}{\widetilde{\chi{}}_{1}^{0}}
\newcommand{\sqa}{\widetilde{q}_1}
\newcommand{\sqb}{\widetilde{q}_2}
\begin{document}
%
%
\title{
Search for the Production of Scalar Bottom Quarks 
in {\boldmath $p\overline{p}$} Collisions at 
{\boldmath $\sqrt{s}$} = 1.96 TeV \\
}

\affiliation{Institute of Physics, Academia Sinica, Taipei, Taiwan 11529, Republic of China} 
\affiliation{Argonne National Laboratory, Argonne, Illinois 60439} 
\affiliation{University of Athens, 157 71 Athens, Greece} 
\affiliation{Institut de Fisica d'Altes Energies, Universitat Autonoma de Barcelona, E-08193, Bellaterra (Barcelona), Spain} 
\affiliation{Baylor University, Waco, Texas  76798} 
\affiliation{Istituto Nazionale di Fisica Nucleare Bologna, $^{cc}$University of Bologna, I-40127 Bologna, Italy} 
\affiliation{Brandeis University, Waltham, Massachusetts 02254} 
\affiliation{University of California, Davis, Davis, California  95616} 
\affiliation{University of California, Los Angeles, Los Angeles, California  90024} 
\affiliation{University of California, San Diego, La Jolla, California  92093} 
\affiliation{University of California, Santa Barbara, Santa Barbara, California 93106} 
\affiliation{Instituto de Fisica de Cantabria, CSIC-University of Cantabria, 39005 Santander, Spain} 
\affiliation{Carnegie Mellon University, Pittsburgh, PA  15213} 
\affiliation{Enrico Fermi Institute, University of Chicago, Chicago, Illinois 60637}
\affiliation{Comenius University, 842 48 Bratislava, Slovakia; Institute of Experimental Physics, 040 01 Kosice, Slovakia} 
\affiliation{Joint Institute for Nuclear Research, RU-141980 Dubna, Russia} 
\affiliation{Duke University, Durham, North Carolina  27708} 
\affiliation{Fermi National Accelerator Laboratory, Batavia, Illinois 60510} 
\affiliation{University of Florida, Gainesville, Florida  32611} 
\affiliation{Laboratori Nazionali di Frascati, Istituto Nazionale di Fisica Nucleare, I-00044 Frascati, Italy} 
\affiliation{University of Geneva, CH-1211 Geneva 4, Switzerland} 
\affiliation{Glasgow University, Glasgow G12 8QQ, United Kingdom} 
\affiliation{Harvard University, Cambridge, Massachusetts 02138} 
\affiliation{Division of High Energy Physics, Department of Physics, University of Helsinki and Helsinki Institute of Physics, FIN-00014, Helsinki, Finland} 
\affiliation{University of Illinois, Urbana, Illinois 61801} 
\affiliation{The Johns Hopkins University, Baltimore, Maryland 21218} 
\affiliation{Institut f\"{u}r Experimentelle Kernphysik, Karlsruhe Institute of Technology, D-76131 Karlsruhe, Germany} 
\affiliation{Center for High Energy Physics: Kyungpook National University, Daegu 702-701, Korea; Seoul National University, Seoul 151-742, Korea; Sungkyunkwan University, Suwon 440-746, Korea; Korea Institute of Science and Technology Information, Daejeon 305-806, Korea; Chonnam National University, Gwangju 500-757, Korea; Chonbuk National University, Jeonju 561-756, Korea} 
\affiliation{Ernest Orlando Lawrence Berkeley National Laboratory, Berkeley, California 94720} 
\affiliation{University of Liverpool, Liverpool L69 7ZE, United Kingdom} 
\affiliation{University College London, London WC1E 6BT, United Kingdom} 
\affiliation{Centro de Investigaciones Energeticas Medioambientales y Tecnologicas, E-28040 Madrid, Spain} 
\affiliation{Massachusetts Institute of Technology, Cambridge, Massachusetts  02139} 
\affiliation{Institute of Particle Physics: McGill University, Montr\'{e}al, Qu\'{e}bec, Canada H3A~2T8; Simon Fraser University, Burnaby, British Columbia, Canada V5A~1S6; University of Toronto, Toronto, Ontario, Canada M5S~1A7; and TRIUMF, Vancouver, British Columbia, Canada V6T~2A3} 
\affiliation{University of Michigan, Ann Arbor, Michigan 48109} 
\affiliation{Michigan State University, East Lansing, Michigan  48824}
\affiliation{Institution for Theoretical and Experimental Physics, ITEP, Moscow 117259, Russia} 
\affiliation{University of New Mexico, Albuquerque, New Mexico 87131} 
\affiliation{Northwestern University, Evanston, Illinois  60208} 
\affiliation{The Ohio State University, Columbus, Ohio  43210} 
\affiliation{Okayama University, Okayama 700-8530, Japan} 
\affiliation{Osaka City University, Osaka 588, Japan} 
\affiliation{University of Oxford, Oxford OX1 3RH, United Kingdom} 
\affiliation{Istituto Nazionale di Fisica Nucleare, Sezione di Padova-Trento, $^{dd}$University of Padova, I-35131 Padova, Italy} 
\affiliation{LPNHE, Universite Pierre et Marie Curie/IN2P3-CNRS, UMR7585, Paris, F-75252 France} 
\affiliation{University of Pennsylvania, Philadelphia, Pennsylvania 19104}
\affiliation{Istituto Nazionale di Fisica Nucleare Pisa, $^{ee}$University of Pisa, $^{ff}$University of Siena and $^{gg}$Scuola Normale Superiore, I-56127 Pisa, Italy} 
\affiliation{University of Pittsburgh, Pittsburgh, Pennsylvania 15260} 
\affiliation{Purdue University, West Lafayette, Indiana 47907} 
\affiliation{University of Rochester, Rochester, New York 14627} 
\affiliation{The Rockefeller University, New York, New York 10021} 
\affiliation{Istituto Nazionale di Fisica Nucleare, Sezione di Roma 1, $^{hh}$Sapienza Universit\`{a} di Roma, I-00185 Roma, Italy} 

\affiliation{Rutgers University, Piscataway, New Jersey 08855} 
\affiliation{Texas A\&M University, College Station, Texas 77843} 
\affiliation{Istituto Nazionale di Fisica Nucleare Trieste/Udine, I-34100 Trieste, $^{ii}$University of Trieste/Udine, I-33100 Udine, Italy} 
\affiliation{University of Tsukuba, Tsukuba, Ibaraki 305, Japan} 
\affiliation{Tufts University, Medford, Massachusetts 02155} 
\affiliation{Waseda University, Tokyo 169, Japan} 
\affiliation{Wayne State University, Detroit, Michigan  48201} 
\affiliation{University of Wisconsin, Madison, Wisconsin 53706} 
\affiliation{Yale University, New Haven, Connecticut 06520} 
\author{T.~Aaltonen}
\affiliation{Division of High Energy Physics, Department of Physics, University of Helsinki and Helsinki Institute of Physics, FIN-00014, Helsinki, Finland}
\author{J.~Adelman}
\affiliation{Enrico Fermi Institute, University of Chicago, Chicago, Illinois 60637}
\author{B.~\'{A}lvarez~Gonz\'{a}lez$^v$}
\affiliation{Instituto de Fisica de Cantabria, CSIC-University of Cantabria, 39005 Santander, Spain}
\author{S.~Amerio$^{dd}$}
\affiliation{Istituto Nazionale di Fisica Nucleare, Sezione di Padova-Trento, $^{dd}$University of Padova, I-35131 Padova, Italy} 

\author{D.~Amidei}
\affiliation{University of Michigan, Ann Arbor, Michigan 48109}
\author{A.~Anastassov}
\affiliation{Northwestern University, Evanston, Illinois  60208}
\author{A.~Annovi}
\affiliation{Laboratori Nazionali di Frascati, Istituto Nazionale di Fisica Nucleare, I-00044 Frascati, Italy}
\author{J.~Antos}
\affiliation{Comenius University, 842 48 Bratislava, Slovakia; Institute of Experimental Physics, 040 01 Kosice, Slovakia}
\author{G.~Apollinari}
\affiliation{Fermi National Accelerator Laboratory, Batavia, Illinois 60510}
\author{A.~Apresyan}
\affiliation{Purdue University, West Lafayette, Indiana 47907}
% ...

\collaboration{CDF Collaboration\footnote{With visitors from $^a$University of Massachusetts Amherst, Amherst, Massachusetts 01003,
$^b$Universiteit Antwerpen, B-2610 Antwerp, Belgium, 
$^c$University of Bristol, Bristol BS8 1TL, United Kingdom,
$^d$Chinese Academy of Sciences, Beijing 100864, China, 
$^e$Istituto Nazionale di Fisica Nucleare, Sezione di Cagliari, 09042 Monserrato (Cagliari), Italy,
$^f$University of California Irvine, Irvine, CA  92697, 
$^g$University of California Santa Cruz, Santa Cruz, CA  95064, 
$^h$Cornell University, Ithaca, NY  14853, 
$^i$University of Cyprus, Nicosia CY-1678, Cyprus, 
$^j$University College Dublin, Dublin 4, Ireland,
$^k$University of Edinburgh, Edinburgh EH9 3JZ, United Kingdom, 
$^l$University of Fukui, Fukui City, Fukui Prefecture, Japan 910-0017
$^m$Kinki University, Higashi-Osaka City, Japan 577-8502
$^n$Universidad Iberoamericana, Mexico D.F., Mexico,
$^o$University of Iowa, Iowa City, IA  52242,
$^p$Kansas State University, Manhattan, KS 66506
$^q$Queen Mary, University of London, London, E1 4NS, England,
$^r$University of Manchester, Manchester M13 9PL, England,
$^s$Muons, Inc., Batavia, IL 60510, 
$^t$Nagasaki Institute of Applied Science, Nagasaki, Japan, 
$^u$University of Notre Dame, Notre Dame, IN 46556,
$^v$University de Oviedo, E-33007 Oviedo, Spain, 
$^w$Texas Tech University, Lubbock, TX  79609, 
$^x$IFIC(CSIC-Universitat de Valencia), 56071 Valencia, Spain,
$^y$Universidad Tecnica Federico Santa Maria, 110v Valparaiso, Chile,
$^z$University of Virginia, Charlottesville, VA  22906
$^{aa}$Bergische Universit\"at Wuppertal, 42097 Wuppertal, Germany,
$^{bb}$Yarmouk University, Irbid 211-63, Jordan
$^{jj}$On leave from J.~Stefan Institute, Ljubljana, Slovenia, 
}}
\noaffiliation


\date{\today}

%
%
\begin{abstract}
We report on a search for direct scalar bottom quark (sbottom) pair production in $p \bar{p}$ collisions at 
$\sqrt{s}=1.96$~TeV, in events with large missing transverse energy and two jets of hadrons in
the final state, where at least one of the jets is required to be identified as originating from a $b$ quark. 
The study uses a CDF Run~II data sample corresponding to 2.65~fb${}^{-1}$ of integrated luminosity.
The data are in agreement with the 
standard model.
In an R-parity conserving minimal supersymmetric scenario, and  assuming that the sbottom decays exclusively 
into a bottom quark and a neutralino,   95$\%$ confidence-level
upper limits on the sbottom pair production cross section of 0.1~pb are obtained. 
For neutralino masses below 70~GeV/$c^2$, sbottom masses up to 230~GeV/$c^2$ are excluded at 95$\%$ confidence level.

\end{abstract}

\maketitle



Supersymmetry (SUSY)~\cite{susy} is an extension of the standard model (SM) that 
naturally solves the hierarchy problem~\cite{hierarchy} and provides a possible candidate for dark matter in the Universe.  
SUSY doubles the SM spectrum of particles by introducing a 
new supersymmetric partner (sparticle) for each particle in the SM. 
In particular, a new scalar field is associated with each left- and right-handed quark state, and
two SUSY mass eigenstates $\sqa$ and $\sqb$ result from the mixing of the scalar fields.
In some SUSY scenarios, for which
the ratio of the vacuum expectation values of the two supersymmetric Higgs fields becomes large, 
a significant mass difference between eigenstates in the sbottom quark sector can occur, 
leading to a rather light sbottom $\sbo$ mass state.  
In a generic minimal supersymmetric extension of the SM (MSSM) and assuming R-parity conservation~\cite{susy},
sparticles are produced in pairs and the lightest supersymmetric 
particle (LSP) is stable and   
identified as  the neutralino $\neu$.
Assuming a SUSY particle
mass hierarchy such that the sbottom decays exclusively as $\sbo \to b \neu$, 
the expected signal for direct sbottom pair production is characterized by the presence 
of  two energetic jets from the hadronization of the bottom quarks and 
large missing transverse energy $\met$~\cite{coord} from the two LSPs  in the final state.
Results on searches in this channel using Tevatron data have been previously 
reported by both the CDF and $\Dzero$ experiments in Run II~\cite{tev-cdf}~\cite{tev-d0}.
This Letter presents new results based
on an almost ten times larger data sample collected by the CDF experiment,  
corresponding to  2.65~$\rm fb^{-1}$ of integrated luminosity.   


The CDF II detector is described in detail elsewhere~\cite{cdfii}.
The detector has a charged particle tracking system that is immersed in a 1.4~T solenoidal magnetic field 
coaxial with the beam line, and provides coverage in the 
pseudorapidity~\cite{coord} range $|\eta| \leq 2$. 
Segmented sampling calorimeters, arranged in a projective tower
geometry, surround the tracking system and measure the energy of 
interacting particles for $|\eta| < 3.6$. 
Cherenkov counters in the region $3.7 < |\eta| < 4.7$ measure the number of 
inelastic $p \overline{p}$ collisions to determine the luminosity~\cite{clc}.



The data are collected using a three-level trigger system that selects events with at 
least one calorimeter cluster  with $E_T$ above 100~GeV. 
The events are then required to have a primary vertex
with a $z$ position within 60~cm of the nominal interaction point. 
Jets are reconstructed from the energy deposits in the calorimeter towers using a cone-based 
algorithm~\cite{jetclu} with cone radius $R=\sqrt{\Delta\eta^2 + \Delta\phi^2 } = 0.4$,
and the measured $\etjet$ is corrected for detector effects and contributions from 
multiple $p\bar{p}$ interactions per crossing at high instantaneous luminosity, as
discussed in Ref.~\cite{jetcornim}. 
The events are initially selected with $\met > 10$~GeV and  
two jets with corrected transverse energy 
$\etjet > 25$~GeV and  pseudorapidity   $|\etajet|<2.0$, where 
at least one of the two jets is required to have  $|\etajet|<1.1$.
Events with additional jets with $\etjet > 15$~GeV and  $|\etajet|<2.8$ are 
rejected.
The effect of the trigger selection is studied using control data samples, parametrized 
as a function of $\met$ and the transverse energy of the leading jet $\etjeta$, and 
consistently taken into account~\cite{gdl}. At a given $\met$, the trigger efficiency 
increases with increasing $\etjeta$. 
Finally, at least one of the two leading jets is required to  
originate from a $b$-quark jet candidate,  
using the default CDF $b$-jet tagging algorithm ({\sc{secvtx}})~\cite{svx}, based on  the presence of a
displaced vertex due to the decay of a $b$ hadron inside the jet.

The SM QCD multijet background contribution with large $\met$, 
due to the mismeasurement of the jet energies in the calorimeters, 
is suppressed by requiring an azimuthal separation $\Delta \phi (\met - \rm jet) > 0.4$ for the 
two jets in the event. The SM background contributions with energetic 
electrons~\cite{charge} from $W$ and $Z$ 
decays and reconstructed as jets in the final state 
are suppressed by requiring $ \etjetem / \etjet < 0.9$ for each jet, 
where $\etjetem$ denotes the electromagnetic component of the jet transverse energy. 
In addition, events 
with isolated tracks with transverse momentum $\pttrk$ above 10~GeV/$c$ are vetoed, thus 
rejecting backgrounds with $W$ or $Z$ bosons decaying into muons or tau leptons. 
Beam-related backgrounds and cosmic rays are eliminated by  
requiring an average jet electromagnetic fraction 
$ f_{\rm em} > 0.15$ and 
 an average charged particle
fraction $ f_{\rm ch} > 0.15$, as defined 
in~\cite{sglu}. The requirements on $f_{\rm em}$ and $ f_{\rm ch}$ reject events with anomalous 
energy deposition in the hadronic section of the calorimeter or energy deposits in the calorimeter 
inconsistent with the observed activity in the tracking system,
 and have no significant effect on SUSY signal and SM background physics samples.
%


The dominant source of background in the analysis is due to
events with a   
light-flavor jet which  
is misidentified as  a $b$-jet (mistags). There are also contributions from heavy-flavor QCD multijet events 
with large $\met$ and passing the  $\Delta \phi (\met - \rm jet)$ requirement discussed above. 
In both cases, the background estimation is extracted from data using a procedure similar to that   
described  in \cite{svx} and \cite{hf},     
where different data-driven multidimensional parameterizations are employed to estimate 
the probability for a light-flavor jet in each event to be mistagged, and the probability for a  heavy-flavor jet 
in QCD multijet events to pass the CDF $b$-tag requirements, respectively.   


Simulated event samples are used to determine detector acceptance and 
reconstruction efficiency, estimate the contribution from the 
rest of the SM backgrounds with heavy-flavor jets in the 
final state, and compute the 
number of expected SUSY signal events.  Samples of simulated $\tt$ and 
diboson ($WW$, $WZ$, and $ZZ$) processes are generated using the {\sc pythia} 6.216~\cite{pythia} 
Monte Carlo generator with Tune~A~\cite{underlying}, and 
normalized to next-to-leading order (NLO) predictions~\cite{topnlo,dibosonnlo}.
Samples of simulated $Z/\gamma^* (\to l^+ l^-)$+jets ($l=e,\mu,\tau)$, $Z (\to \nu \bar{\nu})$+jets, and $W (\to l \nu)$+jets events with light- and heavy-flavor jets are
generated using the {\sc alpgen} 2.1 program~\cite{alpgen} interfaced with the 
parton-shower model from {\sc pythia}.  
The normalization of the boson plus jets heavy-flavor samples
includes an  additional multiplicative factor $k_{HF} = 1.4 \pm 0.4$ which brings the predicted light- to heavy-flavor relative
contributions in agreement with data~\cite{svx}. The complete 
$Z/\gamma^*$+jets and $W$+jets samples are then 
normalized to the measured $Z$ and $W$ inclusive cross sections~\cite{dy}.
Finally, samples of single top  events are produced using the {\sc madevent} generator~\cite{madevent} 
and normalized according to NLO predictions~\cite{stnlo}. 
%
%
The SUSY samples are generated in the framework of MSSM 
using {\sc pythia}. Masses  $\sbo$ and $\neu$ are fixed, and $\sbo$ is only allowed to decay via 
$\sbo \to  b \neu$ channel. 
 A total of 106 different samples 
have been generated with sbottom mass $M_{\sbo}$ in the range 
 between 80~GeV/$c^2$ and 280~GeV/$c^2$ and neutralino mass $M_{\neu}$ up to 100~GeV/$c^2$.
The samples are normalized to  
NLO predictions, as implemented in {\sc prospino2}~\cite{prospino}, using CTEQ6.6~\cite{cteq} parton distribution
functions (PDFs) 
and renormalization and factorization scales set to $M_{\sbo}$.
 The production cross section depends on $M_{\sbo}$ and decreases from  50~pb to 0.01~pb as 
the sbottom mass increases.
The Monte Carlo generated events are passed through a full
CDF II detector simulation (based on  {\sc geant3}~\cite{geant} and {\sc gflash}~\cite{gflash})
and reconstructed 
and analyzed with the same analysis chain as for the data.


An optimization is performed with the aim to maximize the sensitivity to a SUSY signal 
across the $\sbo - \neu$ mass plane. For each of the 106 signal samples considered,  
the procedure  
maximizes $S/\sqrt{B}$, where $S$ denotes the number of SUSY
events and  $B$ is the total SM background. 
As the difference
$\Delta M \equiv M_{\sbo} - M_{\neu}$ increases, 
the optimal thresholds on $\met$,  $\etjeta$, $\etjetb$ and $\Delta \phi (\met - \rm jet(2))$ for the second leading jet, and $H_T$, defined as $ H_T = \sum_{\rm i} \etjeti$ (i=1-2), increase.
The results from the different SUSY samples 
are combined to define two single sets 
of thresholds (see Table~1) 
that maximize the 
search sensitivity 
in the widest range of sbottom and neutralino masses at low 
$\Delta M$ ($\Delta M < 90$~GeV/$c^2$) and 
high~$\Delta M$ ($\Delta M > 90$~GeV/$c^2$), respectively.
As an example, for $M_{\neu} = 70$~GeV/$c^2$ and  $M_{\sbo}$ in the range from 150~GeV/$c^2$ to 250~GeV/$c^2$, values for $S/\sqrt{B}$
between 10 and 2.5 are obtained, corresponding to SUSY selection efficiencies of 3$\%$ to 10$\%$.

\begin{table}[htbp]
\begin{footnotesize}
\begin{center}
\renewcommand{\baselinestretch}{1.2}
\begin{tabular}{|c c c c c c |} \hline
\multicolumn{6}{|c|}{\small{Optimal thresholds}} \\ \hline\hline
                   & $\met$ & $H_T + \met$ & $\etjeta$ & $\etjetb$ & $\Delta \phi (\met - \rm jet(2))$ \\ 
                   &  \multicolumn{4}{c}{GeV}  & radians   \\  
low $\Delta M$     &  60     & 165         &  80       &    25     &  0.7           \\ 
high $\Delta M$    &  80     & 300         &  90       &    40     &  0.7           \\ \hline
 \end{tabular}
\end{center}
\renewcommand{\baselinestretch}{1}
\end{footnotesize}
\label{tab:table1}
\caption{
Optimized thresholds on $\met$, $H_T + \met$, $\etjeta$, $\etjetb$, and  $\Delta \phi (\met - \rm jet(2))$ 
for each $\Delta M$ region.
}
\end{table}


A number of control samples in data is considered to test the validity of the SM background predictions.
Samples dominated by QCD multijet events or mistags are obtained by reversing the requirements on 
either $\met$ or $\etjetb$,  
or using a non-overlapping sample selected with looser $b$-tagging requirements.
Control samples dominated by $Z/\gamma^*$+jets, $W$+jets, and top-quark processes, with highly 
energetic leptons in the final state, are
obtained after requiring either
$ \etjetem / \etjet > 0.9$ 
for at least one of the jets, or selecting events with isolated tracks with $\pttrk > 10$~GeV/$c$. 
Agreement is observed between the data and the SM 
predictions in each of the control samples.


A detailed study of systematic uncertainties was carried out~\cite{gdl}. 
The uncertainty on the SM background predictions is dominated by
the determination of the $b$-jet mistag rates, which propagates into an uncertainty in the SM prediction 
between 13$\%$ and 11$\%$ as $\Delta M$ increases. The uncertainty on 
the $k_{HF}$ value applied to the boson plus jets heavy-flavor samples translates into 
an 11$\%$ uncertainty in the SM predictions. The dependence on the 
amount of initial-state and final-state radiation in the Monte Carlo generated samples
for top, boson plus jets, and diboson contributions introduces a 6$\%$  
uncertainty on the SM predictions. Other sources of uncertainty on the predicted SM background 
are: 
a 3$\%$ uncertainty due to the determination of the $b$-tagging 
efficiency in the Monte Carlo simulated samples; a 3$\%$ uncertainty from   
the uncertainties on the absolute normalization of the top quark, diboson, $W$, and $Z$ Monte Carlo 
generated processes;
and a $2.5 \%$
uncertainty  from the knowledge of the jet energy scale~\cite{jetcornim}.  
In addition, uncertainties related to trigger efficiency and the heavy-flavor QCD multijet background
contribute less than 1$\%$ to the final uncertainty. The total systematic uncertainty 
on the SM predictions varies between 19$\%$ and 18$\%$ as $\Delta M$ increases.
In the case of the MSSM signal, 
various sources of uncertainty on the predicted cross sections at NLO, as determined using {\sc prospino2},  
are considered:
the uncertainty due to PDFs is computed using the Hessian method~\cite{hessian} and 
translates into a 12$\%$ uncertainty on the absolute predictions; variations of the renormalization and factorization scale by a factor of two change the theoretical cross sections by about 26$\%$. 
Uncertainties on the amount of initial- and final-state gluon radiation in the
MSSM Monte Carlo generated samples introduce a 10$\%$ uncertainty on the signal yields. The $3\%$ uncertainty on the 
absolute jet energy scale translates into a 9$\%$ to 14$\%$ uncertainty on the MSSM predictions.
 Other sources of uncertainty include: 
a 4$\%$ uncertainty due to the determination of the $b$-tagging 
efficiency, and a 2$\%$ to 1$\%$ uncertainty due to the uncertainty on the  
trigger efficiency. The total systematic uncertainty on the MSSM signal yields varies between 
30$\%$ and $32 \%$ as $\Delta M$ increases. 
Finally, an additional $6 \%$ uncertainty on the quoted total integrated luminosity is also taken into account in both SM background and SUSY signal predictions.



\begin{figure}
\includegraphics[width=\figsize]{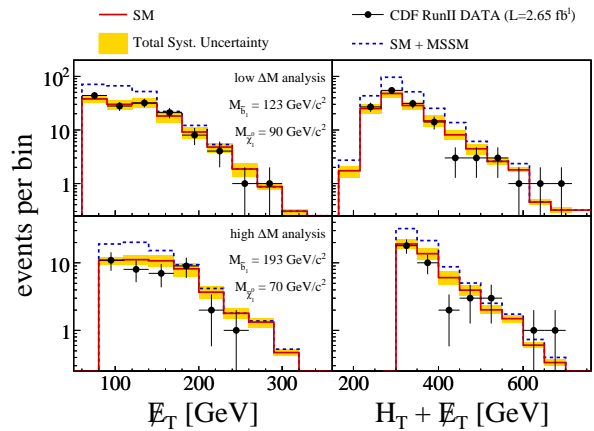}
\caption{ Measured $\met$ and $H_T + \met$ distributions (black dots) for low-$\Delta M$ (top) and high-$\Delta M$ 
analyses (bottom), compared to 
the SM predictions (solid lines) and the SM+MSSM predictions (dashed lines).   
The shaded bands show the total systematic uncertainty on the SM predictions. 
}
\label{fig:metht}
\end{figure}

Figure~\ref{fig:metht} shows the measured $\met$ and $H_T + \met$ distributions 
compared to the SM predictions after all final selection criteria are applied.  
For illustrative purposes, 
the figure indicates the impact of two given MSSM 
scenarios. 
The data are in agreement with the SM predictions within uncertainties
for each of the two analyses at low and high $\Delta M$.
In Table~2, the observed number of events and the SM predictions are presented in each case. 
A global $\chi^2$ test applied to all data points in Fig.~\ref{fig:metht}, and including correlations between systematic uncertainties, gives 
a $30 \%$ probability for data to be consistent with the SM.

\begin{table}[htbp]
\begin{footnotesize}
\begin{center}
\renewcommand{\baselinestretch}{1.2}
\begin{tabular}{|c c c|} \hline
 (2.65 fb${}^{-1}$)                                     & {\small{low $\Delta M$}}  &  {\small{high $\Delta M$}}   \\ \hline
mistags                                &$      51.4\pm{}18.2    $&$     18.5\pm{}5.5     $  \\
QCD jets                               &$      7.6\pm{}1.9    $&$       1.6\pm{}0.2     $  \\
top                                    &$     21.2\pm{}3.4    $&$       7.8\pm{}1.3     $  \\
$Z \rightarrow \nu \bar{\nu}$+jets     &$     27.7\pm{}8.8    $&$      10.9\pm{}3.5     $  \\
$Z/\gamma^* \rightarrow l^+l^-$+jets   &$      0.5\pm{}0.2    $&$      0.11\pm{}0.04     $  \\
$W \rightarrow l \nu$+jets             &$     22.3\pm{}7.3    $&$      7.3\pm{}2.4     $  \\
$WW,WZ,ZZ$                             &$      3.1\pm{}0.5    $&$      1.4\pm{}0.2     $  \\ \hline
SM  prediction               &$       133.8\pm{}26.4      $&$       47.6\pm{}8.8      $  \\ 
Events in data               &        139           &       38 \\ \hline
 \end{tabular}
\end{center}
\renewcommand{\baselinestretch}{1}
\end{footnotesize}
\label{tab:table2}
\caption{
Number of events in data for the two analyses compared to SM  predictions, including statistical and
systematic uncertainties summed in quadrature.
}
\end{table}

\begin{figure}[h]
\includegraphics[width=\figsizes]{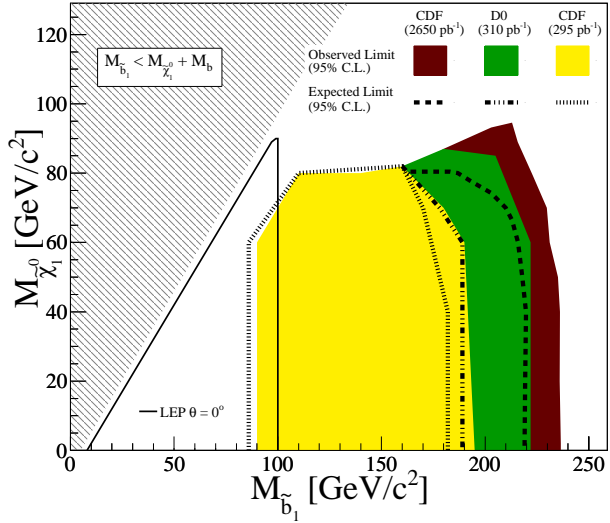}
\caption{
Exclusion plane at 95 $\%$ C.L. as a function of sbottom and neutralino masses.
The observed and expected upper limits from this analysis 
are compared to previous results  from  CDF and $\Dzero$ 
experiments at the Tevatron in Run II~\cite{tev-cdf}~\cite{tev-d0}, and
from LEP~\cite{lep} experiments at CERN with squark mixing angle $\theta = 0^o$.
The hatched area indicates the kinematically prohibited region in the plane.
}
\label{fig:plane}
\end{figure}

The results are translated into 95$\%$ confidence level (C.L.) upper limits on the cross section for 
sbottom pair production at given sbottom and neutralino masses,  
using a Bayesian approach~\cite{bayes} and including
statistical and systematic uncertainties. For the latter, correlations between systematic uncertainties on signal efficiencies 
and background predictions are taken into account. 
For each MSSM point considered, observed and expected limits are computed 
separately for both low- and high-$\Delta M$ 
analyses, and the one with the 
best expected limit is adopted as the nominal result. 
Cross sections above $0.1$ pb are excluded at 
95$\%$ C.L. for the range of sbottom masses considered.
Similarly, the observed numbers of events in data are translated into 95$\%$ C.L. upper limits for sbottom and 
neutralino masses, for which the uncertainties on the theoretical cross sections are also included in 
the limit calculation, and where both analyses are combined in the same way as for the cross section limits. 
Figure~\ref{fig:plane} shows the expected and observed exclusion regions in the sbottom-neutralino mass plane. 
For the MSSM scenario considered, sbottom masses up to 230~GeV/$c^2$ are excluded at 95$\%$ C.L.
for neutralino masses below 70~GeV/$c^2$. This analysis 
extends the previous CDF limits~\cite{tev-cdf} on the sbottom mass by more than 40~GeV/$c^2$.

In summary, we report results on 
a search for sbottom pair production
in  $p \overline{p}$ collisions at $\sqrt{s} = 1.96 \ {\rm TeV}$, 
based on 2.65~fb${}^{-1}$ of CDF Run II data. 
The events are selected with large $\met$ and two energetic jets in the final state, 
and at least one jet is required to originate from a $b$ quark.
The measurements are in agreement with SM predictions for backgrounds. 
The results are translated into 95$\%$ C.L. upper limits on production 
cross sections and sbottom and neutralino masses in a given MSSM scenario
for  which the exclusive decay $\sbo \to b \neu$ is assumed, and 
significantly extend previous CDF results.  


We thank the Fermilab staff and the technical staffs of the participating institutions for their vital contributions. This work was supported by the U.S. Department of Energy and National Science Foundation; the Italian Istituto Nazionale di Fisica Nucleare; the Ministry of Education, Culture, Sports, Science and Technology of Japan; the Natural Sciences and Engineering Research Council of Canada; the National Science Council of the Republic of China; the Swiss National Science Foundation; the A.P. Sloan Foundation; the Bundesministerium f\"ur Bildung und Forschung, Germany; the World Class University Program, the National Research Foundation of Korea; the Science and Technology Facilities Council and the Royal Society, UK; the Institut National de Physique Nucleaire et Physique des Particules/CNRS; the Russian Foundation for Basic Research; the Ministerio de Ciencia e Innovaci\'{o}n, and Programa Consolider-Ingenio 2010, Spain; the Slovak R\&D Agency; and the Academy of Finland. 

%


\begin{thebibliography}{99}
\bibitem{coord} CDF uses a cylindrical coordinate system about the beam axis 
with polar angle $\theta$ and azimuthal angle $\phi$.
We define  transverse energy $E_T = E \ {\rm sin} \theta$, transverse momentum  
$p_T = p \ {\rm sin}  \theta$, pseudorapidity $\eta = - {\rm ln(tan(\frac{\theta}{2})) }$, and rapidity  
$y = \rm \frac{1}{2} ln(\frac{E+p_z}{E-p_z}) $. The missing transverse energy $\met$ is defined as the norm of $-\sum_i E^i_T \cdot \vec{n}_i$, 
where $\vec{n}_i$ is the component in the azimuthal plane of the unit vector pointing
from the interaction point to the i-th calorimeter tower.
\bibitem{charge} Charge conjugation is implied throughout the paper.
\end{thebibliography}
\end{document}